\documentclass[twocolumn,preprintnumbers,superscriptaddress]{revtex4-2}
\usepackage{hyperref}
\usepackage[utf8]{inputenc}

\usepackage{soul}
\usepackage{url}
\usepackage{mathtools}
\usepackage{braket}
\usepackage{amsmath}
\usepackage{dsfont}
\usepackage[english]{babel}
\usepackage{graphicx} 
\usepackage{tikz}
\usetikzlibrary{quantikz}

\usepackage{hyperref}
\usepackage{natbib} 
\usepackage{physics}

\usepackage{listings}
\usepackage{xcolor}

\newcommand{\LDIG}{
Lighthouse Disruptive Innovation Group, LLC
7 Broadway Terrace, Apt 1
Cambridge MA 02139
Middlesex County, Massachusetts (USA)
}

\newcommand{\DSD}{
Engineering Department
Research Group on Data Science for the Digital Society
La Salle - Universitat Ramon Llull
Carrer de Sant Joan de La Salle, 42
08022 Barcelona (Spain)
}
\newcommand{\UVA}{
Universidad de Valladolid
C/Plaza de Santa Cruz, 8, 
47002 Valladolid (Spain)
}

\graphicspath{{./images/}}

\providecommand{\abs}[1]{\lvert#1\rvert}
\providecommand{\norm}[1]{\lVert#1\rVert}

\begin{document}
\title{EVA: a quantum Exponential Value Approximation algorithm}
\author{Guillermo Alonso-Linaje}
\affiliation{\UVA}
\email{guillermo.alonso.alonso-linaje@alumnos.uva.es}

\author{Parfait Atchade-Adelomou}
\affiliation{\DSD}
\email{parfait.atchade@salle.url.edu}
\affiliation{\LDIG}
\email{parfait.atchade@lighthouse-dig.com}
\date{June 2021}

\begin{abstract}
VQE is currently one of the most widely used algorithms for optimizing problems using quantum computers.  A necessary step in this algorithm is calculating the expectation value given a state, which is calculated by decomposing the Hamiltonian into Pauli operators and obtaining this value for each of them. In this work, we have designed an algorithm capable of figuring this value using a single circuit. A time cost study has been carried out, and it has been found that in certain more complex Hamiltonians, it is possible to obtain a good performance over the current methods.
\end{abstract}

\maketitle

\section{Introduction}\label{sec:Introduction}

With the appearance of quantum computing, one of the main fields of application has been optimization problems, which are of great importance in financial, logistics and chemical industry sectors, among others. To solve this type of problem, algorithms such as the Variational Quantum Eigensolver (VQE)\cite {Peruzzo_2014} or Quantum Approximate Optimization Algorithm (QAOA)\cite {farhi2014quantum,Adapt_QAOA}  have emerged whose objective is to find the state that minimizes the expectation value of a given Hamiltonian $ \mathcal {H} $.

Focusing on the VQE, a fundamental stage is the calculation of the expectation value of $ \mathcal {H} $ given an initial state $ \ket {\phi} $. For this, the methodology used consists of decomposing the Hamiltonian as a sum of Pauli operators and then running a circuit for each of these operators to estimate the expectation value of each of them.

As the size of the problems grows, the number of different circuits to run increases dramatically. Due to this fact, various studies have been carried out to reduce the number of circuits, reduce the total amount needed quadratically.

In this paper, we propose a new method with which to be able to approximate the expectation value of $ \mathcal {H} $ with the execution of a single circuit. Finally, we will give an alternative to this algorithm to significantly reduce the depth of the circuit.

After Section \ref{sec:Introduction}, the document is organized as follows; Section \ref{sec:Related_work} shows previous work on both assembly techniques and approaches improve or accelarate VQE; then, Section \ref{sec:Process} presents the quantum fundamentals needed from this era to solve this problem; next, the implementation of the proposed strategy and the creation of the EVA performed in Section \ref{sec:Implementation} are explained; to continue, Section\ref{sec:Error_calculation}, which shows the error calculation of our approach. Next, in Section \ref{sec:Depth_reduction} we present our main strategy for the depth reduction of our approach. Section \ref{sec:result}, which shows the results of our experimental analysis, and Section \ref{sec:Discussions}, in which some open problems are summarized, compared and presented; and finally, Section \ref{sec:Conclusions} concludes the previous results and describes the future work.

\section{Work Context}\label{sec:Related_work}
Variational Quantum Eigensolver (VQE)\cite {Peruzzo_2014} was introduced to reduce the significant hardware demands required by the Phase Estimation Algorithm (PEA)\cite {nielsen2002quantum} and exploit the capabilities of Noisy Intermediate-Scale Quantum devices (NISQ)\cite {Preskill_2018}.

VQE is a hybrid algorithm (classical-quantum) that applies the variational principle of Ritz and that was created for the quantum chemistry of this NISQ era. It was thought that an ansatz would be prepared from a quantum computer that would estimate the expectation value of the electronic Hamiltonian of a molecule. At the same time, a classical optimizer would adjust the parameters of the quantum circuit to find the energy of the ground state of said molecule. Since its publication, the VQE approach has been used in various optimization problems and for quantum chemistry\cite{rattew2019domain, dekeijzer2021optimization, 9248636} and finance \cite{Egger_2020}, logistics\cite{ AtchadeAdelomou2020, atchadeadelomou2021qrobot}, and quantum machine learning\cite {Mar14, JBi17, Adr20, adelomou2020using, atchadeadelomou2021quantum, atchade2021quantum}. Little by little, the VQE has become the flagship of quantum computing, and optimization problems\cite {Grimsley2019}.

Many works of literature have presented techniques and strategies to improve VQE\cite{ wang2019accelerated, Rasmussen2020, Tang_2021, rattew2019domain}. This reference\cite {Rasmussen2020} shows that the number of single-qubit rotations in parameterized quantum circuits can be reduced without compromising the relative expression or entanglement capacity of the circuit. It also shows that the performance of the VQE is not affected by a similar decrease in single-qubit rotations because the relative expression and entanglement capacity are compared at different numbers of qubits in parameterized quantum circuits.

One of the algorithms that has caught our attention and that we have used in previous works is ADAPT-VQE\cite {Tang_2021}.
The ADAPT-VQE was created to decrease the depth of the circuit at the expense of a greater number of measurements. Similar to the original VQE, the ADAPT-VQE algorithm seeks to minimize the circuit's depth further with a greater number of measures.
The algorithm design achieves exact results in convergence. Its approach is based on an ansatz determined by the simulated system and presents an integrated and well-defined convergence criterion. In addition, the parameter count and thus the door depth is kept to a minimum. However, this algorithm is focused on the shallow ansatz and not on the expectation value itself.

Reviewing the state of the art research, we have seen several interesting references\cite {Ekert_2002, Wang_2019, Mitarai_2019, Benfenati2021} that, all and that does not quite solve the execution time problem, present alternatives for the measurement of the expectation value.

When we started the hypothesis of this article last year, we did not see this magnificent work \cite {Verteletskyi_2020} (VQE optimized), which marks an excellent improvement in the reduction of the circuits in Pauli matrices compared to the native VQE and, in the reduction of the execution time. This work, \cite {Verteletskyi_2020} together with the references \cite {jena2019pauli, yen2020measuring, ryabinkin2020iterative, Huggins_2021, gokhale2019minimizing}, have become good tools with which to compare our work due to their great contribution and good performance.
The substantial difference between the \cite {Verteletskyi_2020} approach and those of  \cite {jena2019pauli, yen2020measuring, ryabinkin2020iterative, Huggins_2021, gokhale2019minimizing}, is that the former involves only unit rotations of a qubit, while the latter employ transformations of several qubits.

This work \cite {Mitarai_2019} is the one that is closest to our approach; however, it does not end up solving the execution time problem since its protocols focus on reducing the depth of the quantum circuit significantly by making it unnecessary for the controlled operation to be able to adapt them to the quantum computers of this NISQ era.

In this paper, we propose a new method with which to be able to approximate the expectation value of $ \mathcal {H} $ with the execution of a single circuit, and we compare it with the native VQE and optimized VQE\cite {Verteletskyi_2020}. Finally, we will give an alternative to this algorithm to significantly reduce the depth of the circuit.

\section{The Procedure we follow}\label{sec:Process}

The objective, as we mentioned before, will be to find a method capable of approximate $ \braket {\phi} {\mathcal {H} | \phi} $. However, the Hamiltonian $ \mathcal {H} $ does not have to be unitary, and therefore not constructible in the quantum computer. To solve this problem we will form the complex exponential $ e ^ {i \mathcal {H} t} $.

Once this is done, from a theoretical point of view, we must keep in mind the Taylor expansion of the complex exponential.
\begin{equation}
\label{Taylor_development}
    \braket{\phi}{e^{i\mathcal{H}t}|\phi} = \braket{\phi}{I|\phi} + i\braket{\phi}{\mathcal{H}|\phi}t - \frac{\braket{\phi}{\mathcal{H}^2|\phi}t^2}{2!} -  ...
\end{equation}

On the other hand, we must bear in mind that $ \braket {\phi} {\mathcal {H} ^ m | \phi} $ will be a real number regardless of the state $ \ket {\phi} $ and the value $ m $ chosen. In this way, from \eqref {Taylor_development} we can guarantee the following equality:

\begin{equation*}
\label{parte_imaginaria}
    \Im{\braket{\phi}{e^{i\mathcal{H}t}|\phi}} = \braket{\phi}{\mathcal{H}|\phi}t - \frac{\braket{\phi}{\mathcal{H}^3|\phi}t^3}{3!} +  ...
\end{equation*}

Next, taking a change of variable of $ t $ by $ \frac {1} {k} $ and solving for the previous equation, we get an expression for our expectation value:
\begin{equation*}
\label{previo_fundamental}
    \braket{\phi}{\mathcal{H}|\phi} =  k \Im{\braket{\phi}{e^{\frac{i\mathcal{H}}{k}}|\phi}} + \frac{\braket{\phi}{\mathcal{H}^3|\phi}}{3!k^2} - ...
\end{equation*}

Finally, taking a limit on this result, we arrive at what will be our fundamental equation\eqref{fundamental_result}:
\begin{equation}
\label{fundamental_result}
    \braket{\phi}{\mathcal{H}|\phi} =  \lim_{k \rightarrow \infty}k\Im{\braket{\phi}{e^{\frac{i\mathcal{H}}{k}}|\phi}}
\end{equation}

\section{Implementation}\label{sec:Implementation}

The first step of the algorithm is to construct the complex exponential of the Hamiltonian. This procedure can be carried out through Troterization, giving us a method to approximate the value sought. However, it should be mentioned that the process is exact for the entire set of problems based on the Ising model, that is, those whose Hamiltonian can be expressed as:

\begin{equation}
\label{Ising_model}
   \mathcal{H} = \sum_{i,j} \alpha_{ij} \sigma_i^z \sigma_j^z + \sum_{i}\beta{i} \sigma_i^z 
\end{equation}

where $ \alpha_{ij}, \beta{i} $ real numbers, $ \sigma_i ^ z $ is the tensor product of identity operators except for an operator $ \sigma ^ z $ in the $i-th$ position and in the same way, $ \sigma_i ^ z \sigma_j ^ z $ product of identity operators except for positions $ i $ and $ j $.

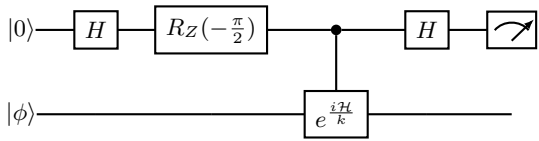
\begin{figure}
\begin{center}
\begin{quantikz}
\ket{0} & \gate{H}{}  & \gate{R_Z(-\frac{\pi}{2})}{}  &\ctrl{1} & \gate{H}{} & \meter{} \\
\ket{\phi} & \qw & \qw & \gate{e^{\frac{i\mathcal{H}}{k}}}{} & \qw & \qw \\
\end{quantikz}
\end{center}
\centering
\caption{Hadamard Test on the exponential gate modified with a $ R_Z $ gate to obtain the imaginary part of the expectation value.}
\centering
\label{fig:img_HTest}
\end{figure}

Throughout this work, we will attack optimization problems whose Hamiltonian is defined in this way. Therefore, we can construct exactly $ e ^ {\frac {i \mathcal {H}} {k}} $ as indicated in Appendix \ref{sec:constr_exp}. Subsequently, a study will be made for the correct choice of $ k $, which we will always assume greater than $ 1 $. At this point, we just have to modify the Hadamard Test capable of returning the imaginary part of an expectation value as we can see in Fig.\eqref{fig:img_HTest}.

In this way, if we measure the first qubit, we arrive at the following equality:

\begin{equation}
\label{Resultado_circuito}
    P(0) - P(1) = \Im{\braket{\phi}{e^{\frac{i\mathcal{H}}{k}}|\phi}}
\end{equation}

Being $ P (0) $ and $ P (1) $ the probability of obtaining $ \ket {0} $ and $ \ket {1} $ when measuring on said qubit. We will leave the development of equality in the Appendix \eqref{sec:HTest_A}. Therefore, multiplying by $ k $ based on \eqref {fundamental_result} we obtain our approximation to $ \bra {\phi} \mathcal {H} \ket {\phi} $.

\section{Error calculation} \label{sec:Error_calculation}

For this section, we will assume that we have performed a previous normalization of the Hamiltonian, that is:

\begin{equation*}
    \norm{\mathcal{H}} := \text{max}_{\phi} |\bra{\phi}\mathcal{H}\ket{\phi}| \leq 1
\end{equation*}

First of all, we must see how well our method approximates the real value as a function of $ k $; therefore, using \eqref{previo_fundamental}, we obtain the equation \ref{error}.

\begin{equation}
    \label{error}
    \abs{\Im{\braket{\phi}{e^{\frac{i\mathcal{H}}{k}}|\phi}} - \frac{\braket{\phi}{\mathcal{H}|\phi}}{k}} \leq \frac{\norm{\mathcal{H}}^3}{6k^3}
\end{equation}

\begin{figure}[!ht]
    \centering
    \includegraphics [width=0.4\textwidth]{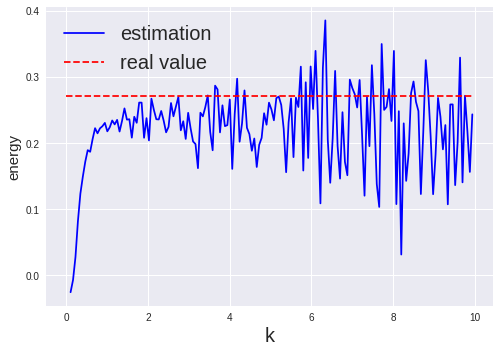}
    \caption{In this image, we can see the behaviour of the model when estimating the expectation value $ \braket {\phi} {\mathcal {H} | \phi} $ set several shots. It quickly converges to the solution, but as the value of k continues to advance, the uncertainty becomes greater.}
    \label{fig:k_error}
\end{figure}

We will leave the development of this inequality in the Appendix \eqref{sec:Err_Formula}. However, from the equation \eqref{error}, we obtain very valuable information: firstly, we show that as $ k $ grows, the distance between both values decreases and secondly; when the distance between all expectation values decreases (a consequence of having $ \norm {\mathcal {H}} $ small), the model becomes more accurate.
It may seem that, in this way, it would be better to choose a value of $ k $ as large as possible; however, as we can see in Fig.\eqref{fig:k_error}, setting the number of shots the model does not behave in this way, in the long run, the deviation increases.

This is because we are not trying to estimate $ \bra {\phi} \mathcal {H} \ket {\phi} $ otherwise $ \frac {\bra {\phi} \mathcal {H} \ket {\phi }} {k} $, and to approximate those values with the same precision we would have to make $ k ^ 2 $ times more shots in the second model. 

\begin{figure}[!ht]
    \centering
    \includegraphics [width=0.4\textwidth]{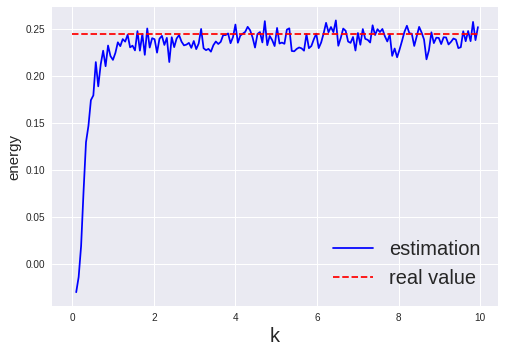}
    \caption{In this case, with the number of shots dependent on $ k $, the behavior of the model when estimating the expectation value $ \braket {\phi} {\mathcal {H} | \phi} $ is stable as $ k $ advances.}
    \label{fig:k_error2}
\end{figure}

As can be seen in Fig.\eqref{fig:k_error2}, regulating the number of shots, the variance concerning the expectation value will remain. From here, you can understand the importance of the parameter $ k $ in the algorithm since it plays two very important roles: establishing the accuracy of the method concerning the expectation value of $ \mathcal {H} $ (the larger $ k $, the more accurate ) and condition the number of shots that will be needed (the smaller $ k $, the fewer shots will be required).
We consider it important to mention that the growth of the number of shots is quadratic while the error decreases in a cubic way, thus controlling that the number of shots will not end up firing as the precision needs to be increased.
Empirically it has been seen that for Hamiltonians with many terms, values between $ 2 $ and $ 4 $ are good approximations for $ k $. Also as the problems grow, the normalizations of $ \mathcal {H} $ get tougher by making $ \norm {\mathcal {H}} $ smaller by decreasing the error in \eqref{error}.

\section{Depth reduction} \label{sec:Depth_reduction}

Although EVA currently presented has a great performance, as we will see later in the results, this is obtained through an expensive circuit. The reason for this is that we are controlling the exponential matrix, which is made up of a large number of C-NOT gates that will eventually transform into Toffoli gates. To build one of these gates, you need at least six C-NOTs to increase the depth of the circuit. For this reason, we propose an algorithm capable of approximating the solution without any Toffoli gate, the reduced EVA. The new proposed structure would be the following Fig.\eqref{fig:gTv2}
\begin{figure}
\begin{center}
\begin{quantikz}
\ket{0}    & \gate{H}{}  & \gate{R_Z(\frac{\pi}{2})}{}  &\ctrl{1} & \qw &\ctrl{1} & \gate{H}{} & \qw\\
\ket{\phi} & \qw         & \qw                          & \gate{H}{} &\gate{e^{\frac{i\mathcal{H}}{k}}}{} &\gate{H}{} & \qw & \qw \\
\end{quantikz}
\end{center}
\centering
\caption{Modification of the Hadamard Test to decrease the depth of the circuit. This new circuit approximates the solution under specific conditions in the text.}
\centering
\label{fig:gTv2}
\end{figure}
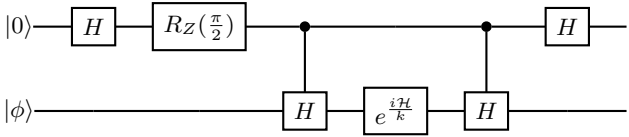
where  state $ \ket {\phi} $ can occupy more than one qubit. In this circuit, a Control-H gate would be made for each of these qubits.
Advancing the idea, it is important to note that both circuits are not equivalent, but they are under specific conditions that already satisfy our problem:

\begin{itemize}
    \item The angles of the exponential must be small: this is something that we had to guarantee to make the approximation \eqref{fundamental_result}.
    \item The ansatz must be defined in a single axis: when working with optimization problems on $Z$ we can fulfil this property without difficulty.
\end{itemize}

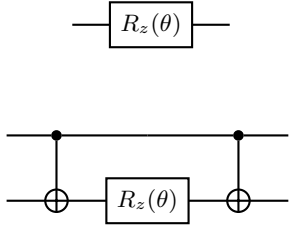
\begin{figure}
\begin{center}
\begin{quantikz}
\\
\qw & \gate{R_z(\theta)}{} & \qw\\
\end{quantikz}
\end{center}

\begin{center}
\begin{quantikz}
\qw &\ctrl{1}  & \qw                        & \ctrl{1} & \qw \\
\qw &\targ{} & \gate{R_z(\theta)}{} & \targ{} & \qw
\end{quantikz}
\end{center}
\centering
\caption{With the first circuit we can construct the exponential of $ \sigma_i $ and with the second, the exponential of $ \sigma_i \sigma_j $.}
\centering
\label{fig:Trottetization_gT}
\end{figure}

By the definition of $ e ^ {i \mathcal {H}} $, we will only find blocks of two types Fig.\eqref{fig:Trottetization_gT}.
It would be enough to prove that the equality of circuits is fulfilled for each of these two blocks since for a larger Hamiltonian, it would only be necessary to concatenate them. However, in this paper, we will only show the development for $ e ^ {i \mathcal {H}} = R_z (\theta) $ since the calculations for the other are performed analogously.

The first step will be to determine what is the goal you want to reach. To do this, given a state $ \ket {\phi} = \alpha \ket {0} + \beta \ket {1} $, we get that:

\begin{equation}
    \label{im_gt1}
    \Im( \bra{\phi} R_z(\theta) \ket{\phi}) = (-|\alpha|^2 + |\beta|^2)\sin{\frac{\theta}{2}} 
\end{equation}

The idea, therefore, will be to check if we arrive at the same solution through our new approach \eqref{fig:gTv2}. Initially, we find the state $ \ket{\Phi_0} \coloneqq \ket{0} \ket{\phi} $ which after applying the first two gates on the first qubit will remain as:
\begin{equation*}
    \ket{\Phi_1} \coloneqq \frac{1}{\sqrt{2}}(\ket{0}\ket{\phi}+i\ket{1}\ket{\phi})
\end{equation*}

Subsequently applying the Hadamard gates only to the qubit that has control at $ 1 $ between $ R_z $ we arrive at the following expression:

\begin{equation*}
    \ket{\Phi_2} \coloneqq \frac{1}{\sqrt{2}}(\ket{0}R_z(\theta)\ket{\phi}+i\ket{1}HR_z(\theta)H\ket{\phi})
\end{equation*}

Finally, remembering that $ HR_zH = R_x $ and applying the last Hadamard gate, we obtain our final state as:

\begin{equation*}
    \ket{\Phi_3} \coloneqq \frac{1}{2}(\ket{0}(R_z(\theta)+iR_x(\theta))\ket{\phi}+\ket{1}(R_z(\theta)-iR_x(\theta))\ket{\phi})
\end{equation*}

We can already calculate the expectation value on the first qubit from this final state, that is, $P(0) -P(1) $. The following is the result for which $ \beta $ is assumed to be of the form $ i \beta '$ where $ \beta' $ is a real number:

\begin{equation*}
    P(0)-P(1) = \frac{(-|\alpha|^2 + |\beta|^2)\sin{\theta}}{2} - \alpha\beta'(1-\cos{\theta})
\end{equation*}

We will leave the development up to this equality in the Appendix \eqref{sec:Expect_Val_A}. This expression is not comparable with the goal set in \eqref{im_gt1}, but we can use a hypothesis that we were considering: the values of $ \theta $ are controlled to be small. In this way, applying infinitesimals, we obtain:

\begin{equation*}
    P(0)-P(1) = (-|\alpha|^2 + |\beta|^2)\sin{\frac{\theta}{2}}  + \alpha\beta'\frac{\theta^2}{2}
\end{equation*}

As we are working with low values, we can assume that $ \theta ^ 2 \ll \theta $ so the last term is neglected reaching the desired equality. The consequence of taking $ \beta $ in this way forces us that when creating our ansatz, they can only be made through $ R_x $ and CNOT gates (which is not an issue for solving problems in a single axis).
As we have just verified, we are working with an approximation method, so the first approximation error will have to be added to the accumulated by this phase. Operating with this new error, we get that:

\begin{equation*}
    \left|P(0)-P(1) - \frac{\bra{\phi}\mathcal{H}\ket{\phi}}{k}\right| \leq \frac{\norm{\mathcal{H}}^2}{2k^2} 
\end{equation*}

Given this, the error and the number of shots used become inversely proportional magnitudes. Remember anyway that as the size of the problems grows, the normalizations of the Hamiltonian cause that $ \norm {\mathcal {H}} ^ 2 $ regulates the error itself.

\section{Results}\label{sec:result}

A comparative study of the efficiency (in terms of time) of the two versions of EVA, the VQE and the optimized VQE \cite{Verteletskyi_2020}, has been carried out. To do this, it has been decided to create a series of random Hamiltonians according to the equations \eqref{ising_eq} and \eqref{huso}.

\begin{equation}
\label{ising_eq}
   \mathcal{H} = \sum_{i,j} \alpha_{ij} \sigma_i^z \sigma_j^z + \sum_{i}\beta{i} \sigma_i^z 
\end{equation}

\begin{equation}
\label{huso}
   \mathcal{H} = \sum_{i,j,k} \delta_{ijk} \sigma_i^z \sigma_j^z \sigma_k^z+ \sum_{i,j} \alpha_{ij} \sigma_i^z \sigma_j^z + \sum_{i}\beta_{i} \sigma_i^z 
\end{equation}

where the coefficients will take the value $ 0 $ with a certain probability $ p $, which will determine the size of the Hamiltonian. Unlike computers that solve QUBO-type problems through annealing, optimization through gates allows working with polynomials of degree greater than two without adding any auxiliary variable. This is the example of the Hamiltonian defined as in \eqref{huso}, which it has been decided to include in the study since the behaviours of the algorithms through higher degree Hamiltonians vary considerably.

\begin{figure}[!ht]
\label{img1}
    \centering
    \includegraphics [width=0.4\textwidth]{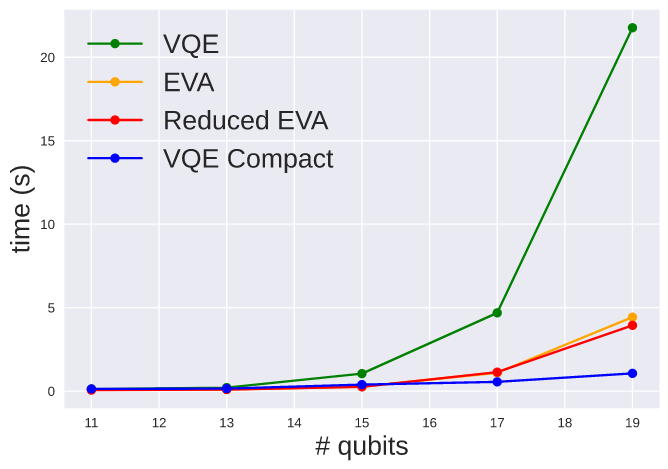}
    \caption{In this figure, we can appreciate the growth in a time of the different algorithms, in this case for $p = 0.5$ and taking Hamiltonians of the Ising form (degree 2). In green, we show the VQE; in blue its optimized version\cite{Verteletskyi_2020}, in yellow the EVA and in red the reduced EVA.}
\end{figure}

\begin{figure}[!ht]
\label{img2}
    \centering
    \includegraphics [width=0.4\textwidth]{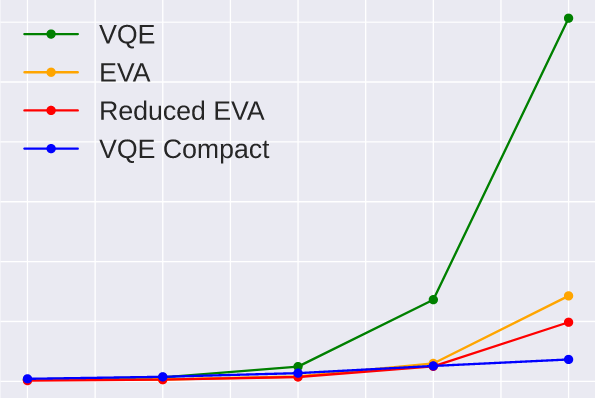}
    \caption{In this second graph for $p = 0.75$ and taking the degree = 2, we can begin to notice a somewhat more significant difference between the first version of the EVA and its improved version. However, the optimized VQE \cite{Verteletskyi_2020} still shows a great advantage over the rest of the algorithms.}
\end{figure}

As you can see from the graphs, the optimized version gets really good performances when working with quadratic problems, while the non-optimized VQE quickly shoots up.

\begin{figure}[!ht]
\label{img3}
    \centering
    \includegraphics [width=0.4\textwidth]{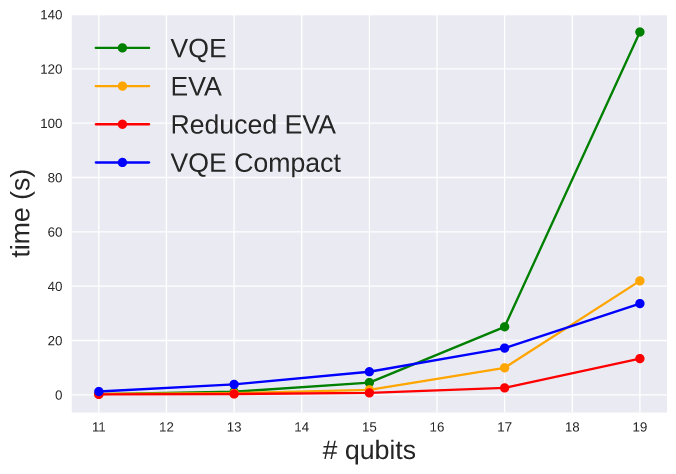}
    \caption{Then, going up to Grade 3 Hamiltonians with $p = 0.5$, we can see that the behaviours are significantly different. At this point, it can be seen how the second version of the EVA (in red) already manages to outperform the rest of the algorithms (VQE, optimized VQE\cite{Verteletskyi_2020}) shown in time efficiency.}
\end{figure}

\begin{figure}[!ht]
\label{img4}
    \centering
    \includegraphics [width=0.4\textwidth]{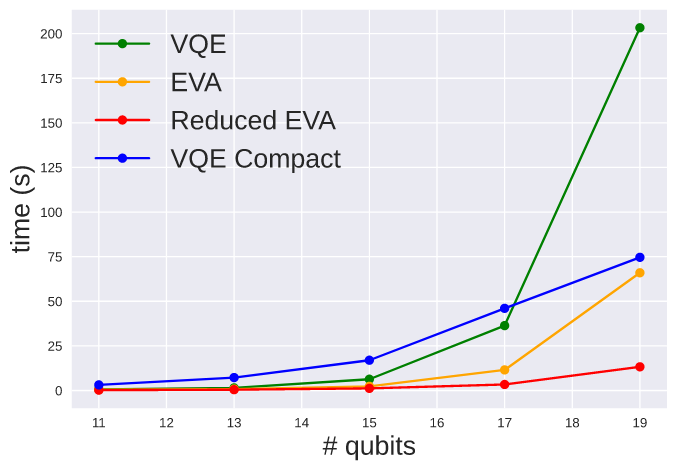}
    \caption{In this last comparison, in which grade = 3 and $p = 0.75$ are being taken, we can observe a very similar behaviour to the previous situation. However, although in this case, the first version of EVA seems to outperform the the optimized VQE \cite{Verteletskyi_2020}, it is clear that the trend, in the long run, will be significantly worse.}
\end{figure}

However, we can see that the optimized version of VQE is no longer as efficient. At this point, it can be seen that although the trend of the first version of the EVA will end up being worse, the improved performance of it will be much more competitive. To reproduce all the scenarios, you will find the code in python in this Ref.\cite{EVA_code}.

\section{Discussions}\label{sec:Discussions}

Throughout this study, we have seen developing a new technique for calculating the expectation value of a Hamiltonian. Although, unlike other traditional methods, we are carrying out an approximation algorithm, the results have been very positive, reaching very precise estimates, drastically reducing the time. However, it is important to note that at the simulation limit, reaching $ 23 $ qubit, a change in trend has been observed between the optimized VQE and the second version of the EVA. This change in trend could be due to the operation of the simulator itself since the number of matrices multiplied in our algorithm is much higher due to the greater number of gates used, and these operations would not be carried out in real hardware. We will have to wait for the availability of new quantum computers to truly verify the scalability of the algorithm for a number of qubits greater than $ 23 $.

\begin{figure}[!ht]
\label{img4}
    \centering
    \includegraphics [width=0.4\textwidth]{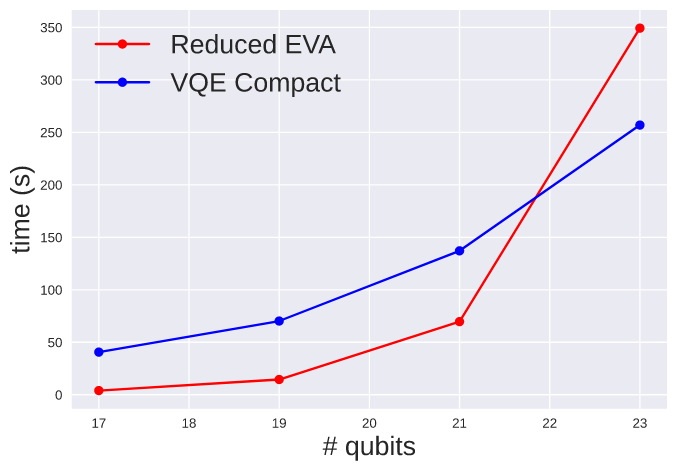}
    \caption{This graph, focusing only on the Reduced EVA in red and optimized VQE\cite{Verteletskyi_2020} in blue, shows a change in trend over time working with the simulator offered by Pennylane.}
\end{figure}

\section{Conclusions and further work}\label{sec:Conclusions}

The EVA has proven to be a good method for calculating expectation Hamiltonian values for optimization problems. Furthermore, it begins to gain strength with issues in which the degree of the variables is greater than two.
It can be seen as a good starting point to start working with more complex Hamiltonians playing with Pauli gates in various directions to apply to a much higher range of problems.

In future work, we will study how the error produced by Troterization affects the algorithm when the Hamiltonian does not follow the Ising model, and a more in-depth development will be made on its usefulness in optimization algorithms in real hardware.

\acknowledgments

The authors greatly thank the Pennylane team, especially Joshua Izaac, for their active support throughout the project and for facilitating its development.

\appendix \label{sec:Appendix}

\section{Construction of the exponential}\label{sec:constr_exp}

The goal will be $ e ^ {i \mathcal {H} t} $ that being a complex exponential, and we know that it is unitary and therefore can be constructed in a quantum circuit. We can decompose the Hamiltonian by its definition in \eqref {Ising_model}, arriving at \eqref{exp_H}.

\begin{equation}
\label{exp_H}
    e^{i\mathcal{H}t} = (\prod_{k,j}e^{i\alpha_{kj}\sigma_k^z \sigma_j^z t})(\prod_{j}e^{i\beta_{j} \sigma_j^z t})
\end{equation}

That is, you just have to define the circuit that forms each of the exponentials separately and concatenate them one after the other. When working with an Ising model, this procedure can be carried out exactly since the $ Z $ gates are commutative.

For the exponential $ e ^ {i \beta{j} \sigma_j ^ z t} $ we would have to build the circuit:
\begin{figure}
\begin{center}
\begin{quantikz}
\\
\qw & \gate{R_z(-2t\beta_{j})}{} & \qw\\
\end{quantikz}
\caption{Rotation associated with the term $ \sigma_{j} $ whose coefficient is $ \beta_{j} $}
\end{center}

applied this gate to the $j-th$ qubit and with respect to the term $ e ^ {i \alpha_ {kj} \sigma_k ^ z \sigma_j ^ z t} $ the associated circuit would be the following:

\begin{center}
\begin{quantikz}
\qw &\ctrl{1}  & \qw                        & \ctrl{1} & \qw \\
\qw &\targ{} & \gate{R_z(-2t\alpha_{kj})}{} & \targ{} & \qw
\end{quantikz}
\end{center}
\centering
\caption{Rotation associated with the term $ \sigma_{kj} $ whose coefficient is $ \alpha_{kj} $}
\centering
\label{fig:Hadamard_Test2}
\end{figure}
where the qubits involved are the $jth$ and $kth$. 

\section{Hadamard Test}\label{sec:HTest_A}

This section will show why Hadamard Test, where the qubits involved are the $jth$ and $ktht$ with that gate $ R_Z $ a higher, can return the imaginary part of the expectation value. In this case, we will demonstrate it for a generic $ U $ gate:
\begin{figure}
\begin{center}
\begin{quantikz}
\ket{0} & \gate{H}{}  & \gate{R_Z(-\frac{\pi}{2})}{}  &\ctrl{1} & \gate{H}{} & \meter{} \\
\ket{\phi} & \qw & \qw & \gate{U}{} & \qw & \qw \\
\end{quantikz}
\end{center}
\centering
\caption{Generic structure to apply the Hadamard test to an arbitrary gate $U$ with modification ($ R_Z $) to obtain the imaginary part of the expectation value.}
\centering
\label{fig:Hadamard_Test2}
\end{figure}
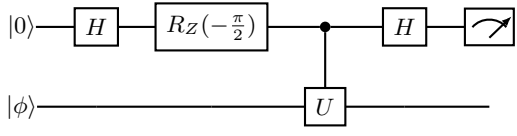

Initially, we start with the state $ \ket {\phi} \ket {0} $, which, after applying the gate $ H $ on the first qubit and subsequently $ R_Z $, will remain as

\begin{gather*}
    \ket{\phi}\frac{\ket{0} - i\ket{1}}{\sqrt{2}}
\end{gather*}

After performing the controlled gate, the status will become:

\begin{gather*}
    \frac{(-iU\ket{\phi})\ket{1}}{\sqrt{2}} + \frac{(\ket{\phi})\ket{0}}{\sqrt{2}} 
\end{gather*}

and by reapplying a gate $ H $ on the first qubit we arrive at the final state:

\begin{gather*}
    \frac{(-iU\ket{\phi})\ket{0} + (iU\ket{\phi})\ket{1}}{2} + \frac{(\ket{\phi})\ket{0} + (\ket{\phi})\ket{1}}{2} 
\end{gather*}

Grouping concerning to the measurement qubit would be as follows:

\begin{gather*}
    \frac{(-iU\ket{\phi} + \ket{\phi})\ket{0} + (iU\ket{\phi}+\ket{\phi})\ket{1}}{2} 
\end{gather*}

Looking at it this way, we can already calculate the probability of obtaining $ 0 $ and $ 1 $ concerning the first qubit by taking modules of their coefficients, that is:

\begin{gather*}
    P(0) = \norm{\frac{(-iU\ket{\phi} + \ket{\phi})}{2}}^2 = \frac{(\bra{\phi} iU^\dag + \bra{\phi})(-iU\ket{\phi} + \ket{\phi})}{4} 
    \\ \\
   = \frac{2 + \bra{\phi} iU^\dag \ket{\phi} - \bra{\phi} iU \ket{\phi} }{4}
\end{gather*}

Analogously, we can calculate the probability of obtaining $ 1 $ as:

\begin{gather*}
    P(1) = \norm{\frac{iU\ket{\phi} + \ket{\phi})}{2}}^2 = \frac{(-\bra{\phi} iU^\dag + \bra{\phi})(iU\ket{\phi} + \ket{\phi})}{4} = 
    \\ \\
   = \frac{2 - \bra{\phi} iU^\dag \ket{\phi} + \bra{\phi} iU \ket{\phi} }{4}
\end{gather*}

And subtracting both probabilities, we can see that:

\begin{gather}
    P(0) - P(1) = \frac{\bra{\phi} iU^\dag \ket{\phi} - \bra{\phi} iU \ket{\phi} }{2}
\end{gather}

Finally, it would only be necessary to realize that if $ z $ is a complex number, it is true that:

\begin{gather}
    \frac{i\hat{z} - iz}{2} = \Im{z}
\end{gather}

We can conclude by saying then that:

\begin{gather*}
    P(0) - P(1) = \Im (U)
\end{gather*}

just as we wanted to demonstrate.

\section{Error formula}\label{sec:Err_Formula}

Let's prove the equation \eqref{error}, the formula for the error of our model. We will start with the expansion of the imaginary part of $ e ^ {i \frac {\mathcal {H}} {k}} $:

\begin{equation*}
    \Im{\braket{\phi}{e^{i\frac{\mathcal{H}}{k}}|\phi}} = \frac{\braket{\phi}{\mathcal{H}|\phi}}{k} - \frac{\braket{\phi}{\mathcal{H}^3|\phi}}{k^3 3!} +  ...
\end{equation*}

Clearing from here and taking modules, we get to:

\begin{equation*}
    \abs{\Im{\braket{\phi}{e^{i\frac{\mathcal{H}}{k}}|\phi}} - \frac{\braket{\phi}{\mathcal{H}|\phi}}{k}} =  \abs{\frac{\braket{\phi}{\mathcal{H}^3|\phi}}{k^3 3!} - \frac{\braket{\phi}{\mathcal{H}^5|\phi}}{k^5 5!} + ...} 
\end{equation*}

As $ \mathcal {H} $ is a Hamiltonian of type \eqref {Ising_model}, it will be a diagonal matrix, and therefore we can guarantee that:

\begin{equation*}
    \braket{\phi}{\mathcal{H}^m|\phi} = \braket{\phi}{\mathcal{H}|\phi}^m 
\end{equation*}

This means that we can see the last previous sequence as a power expansion of $ \frac {\norm {\mathcal {H}}} {k} $ and rebuild to arrive at:

\begin{equation*}
   \abs{\frac{\braket{\phi}{\mathcal{H}^3|\phi}}{k^3 3!} - \frac{\braket{\phi}{\mathcal{H}^5|\phi}}{k^5 5!} + ...} = \abs{\frac{\braket{\phi}{\mathcal{H}|\phi}}{k} - \sin{\frac{\braket{\phi}{\mathcal{H}|\phi}}{k}}}
\end{equation*}

Let $ \norm {\mathcal {H}}: = \text{max}_{\phi} \braket {\phi} {\mathcal {H} | \phi} $ be the norm of $ \mathcal {H} $ , as the function $ f (x) = x- \sin {x} $ increasing and positive (for $ x \in (0,1) $), we can then limit the previous equation by the norm:

\begin{equation*}
   \abs{\frac{\braket{\phi}{\mathcal{H}|\phi}}{k} - \sin{\frac{\braket{\phi}{\mathcal{H}|\phi}}{k}}} \leq \frac{\norm{\mathcal{H}}}{k} - \sin{\frac{\norm{\mathcal{H}}}{k}}
\end{equation*}

and doing the development again on the second term of the inequality, we obtain that:

\begin{equation}
\label{eq_error}
   \abs{\frac{\braket{\phi}{\mathcal{H}^3|\phi}}{k^3 3!} - \frac{\braket{\phi}{\mathcal{H}^5|\phi}}{k^5 5!} + ...} \leq \frac{\norm{\mathcal{H}}^3}{k^3 3!} - \frac{\norm{\mathcal{H}}^5}{k^5 5!} + ...
\end{equation}

Now, focusing on this last term, it is important to realize that:

\begin{equation*}
 \frac{\norm{\mathcal{H}}^5}{k^5 5!} - \frac{\norm{\mathcal{H}}^7}{k^7 7!} + ... \geq 0
\end{equation*}

The proof of this is based solely on taking the terms two by two and seeing that their sum is always positive. This would be enough since the series is convergent. It is easy to check that:

\begin{equation*}
 \frac{\norm{\mathcal{H}}^{4n+1}}{k^{4n+1} (4n+1)!} - \frac{\norm{\mathcal{H}}^{4n+3}}{k^{4n+3} (4n+3)!} \geq 0 
\end{equation*}

since simplifying the expression, it would be equivalent to seeing the following:

\begin{equation*}
 1 - \frac{\norm{\mathcal{H}}^2}{k^2 (4n+3)(4n+2)} \geq 0 
\end{equation*}

and, since $ \frac {\norm {\mathcal {H}}} {k} \leq 1 $ being $ \norm {\mathcal {H}} \leq 1 $ and $ k \geq 1 $, it remains proved the above inequality.
So to conclude, going back to \eqref{eq_error}, since the term on the right is positive and we are subtracting $ \frac {\norm {\mathcal {H}} ^ 3} {k ^ 3 3!} $ From $ \frac {\norm {\mathcal {H}} ^ 3} {k ^ 3 3!} $ a positive quantity, we can limit by:

\begin{equation*}
   \abs{\frac{\braket{\phi}{\mathcal{H}^3|\phi}}{k^3 3!} - \frac{\braket{\phi}{\mathcal{H}^5|\phi}}{k^5 5!} + ...} \leq \frac{\norm{\mathcal{H}}^3}{6k^3 }
\end{equation*}

thus arriving at the desired inequality.

\section{Expectation value in a simplified version}\label{sec:Expect_Val_A}

Our goal will be, given the state:

\begin{equation*}
    \ket{\Phi_3} \coloneqq \frac{1}{2}(\ket{0}(R_z(\theta)+iR_x(\theta))\ket{\phi}+\ket{1}(R_z(\theta)-iR_x(\theta))\ket{\phi})
\end{equation*}

determine the value $ P (0) -P (1) $. To do this, first of all, let us realize that it is equivalent to obtaining $ 1-2P (1) $ since the sum of the probabilities will be equal to $ 1 $. Therefore, focusing on the probability of observing $ \ket {1} $ we know that:

\begin{equation*}
    P(1) = \frac{1}{4}\bra{\phi}(R_z^\dagger(\theta)+iR_x^\dagger(\theta))(R_z(\theta)-iR_x(\theta))\ket{\phi})
\end{equation*}

Operating this expression we get that:

\begin{equation*}
    P(1) = \frac{1}{4}\bra{\phi} 2I - iR_z^\dagger(\theta)R_x(\theta)+iR_x^\dagger(\theta)R_z(\theta)\ket{\phi})
\end{equation*}

Knowing the matrix of $ R_z $ and $ R_x $ on the one hand and taking $ \ket {\phi} \coloneqq \alpha \ket {0} + \beta \ket {1} $ we obtain the expression:

\begin{equation*}
\begin{split}
        P(1) = \frac{1}{4}(2|\alpha|^2+2|\beta|^2+ \sin{\theta}|\alpha|^2-\sin{\theta}|\beta|^2 + \\
     \alpha^*\beta(i\cos{\theta}-\sin{\theta}-i) + \alpha\beta^*(-i\cos{\theta}-\sin{\theta}+i))
\end{split}
\end{equation*}

The choice to take $ \beta = i \beta '$ with real $ \beta' $ at this point is to simplify this equation so that:

\begin{equation*}
\begin{split}
        P(1) = \frac{1}{4}(2+ \sin{\theta}|\alpha|^2-\sin{\theta}|\beta|^2 +
     2\alpha\beta'(-\cos{\theta}+1)
\end{split}
\end{equation*}

Therefore, we finally get to:
\begin{equation*}
    P(0) - P(1) = 1-2P(1) = \frac{(-|\alpha|^2 + |\beta|^2)\sin{\theta}}{2} + \alpha\beta'(\cos{\theta}-1)
\end{equation*}
\bibliography{main}
\end{document}